\pdfoutput=1
\documentclass[aps,prl,twocolumn,reprint,superscriptaddress]{revtex4-1}
\usepackage{graphicx}% Include figure files
\usepackage{dcolumn}% Align table columns on decimal point
\usepackage{natbib,aas_macros}
\usepackage{color}

\usepackage{amsmath}
\usepackage{subfig}
\usepackage{mathtools}

\begin{document}
\title{New Constraints on the Origin of Medium-Energy Neutrinos Observed by IceCube}

\author{Antonio Capanema}
\email{antoniogalvao@aluno.puc-rio.br}
\author{Arman Esmaili}
\email{arman@puc-rio.br}
\affiliation{Departamento de F\'isica, Pontif\'icia Universidade Cat\'olica do Rio de Janeiro, Rio de Janeiro 22452-970, Brazil}
\author{Kohta Murase}
\email{murase@psu.edu}
\affiliation{Department of Physics; Department of Astronomy \& Astrophysics; Center for Particle and Gravitational Astrophysics, The Pennsylvania State University, University Park, Pennsylvania 16802, USA}
\affiliation{Center for Gravitational Physics, Yukawa Institute for Theoretical Physics, Kyoto, Kyoto 606-8502, Japan}

\date{\today}

\begin{abstract}
The recent IceCube publication claims the observation of cosmic neutrinos with energies down to $\sim10$ TeV, reinforcing the growing evidence that the neutrino flux in the 10-100 TeV range is unexpectedly large. Any conceivable source of these neutrinos must also produce a $\gamma$-ray flux which degrades in energy en route to the Earth and contributes to the extragalactic $\gamma$-ray background measured by the Fermi satellite. In a quantitative multimessenger analysis, featuring minimalistic assumptions, we find a $\gtrsim3\sigma$ tension in the data, reaching $\sim5\sigma$ for cosmic neutrinos extended down to $\sim1$~TeV, interpreted as evidence for a population of hidden cosmic-ray accelerators.

%Any conceivable scenario for the origin of high-energy cosmic neutrinos, observed by the IceCube neutrino detector at the South Pole, predicts the generation of accompanied high-energy $\gamma$ rays. 
%Propagation of high-energy photons over cosmological distances initiates an electromagnetic cascade that degrades their energy to $\lesssim1$~TeV energies, contributing to the extragalactic $\gamma$-ray background (EGB) observed by the {\it Fermi-LAT} experiment. 
%By taking into account various established components of the EGB, as well as the latest IceCube shower data in the 10-100~TeV range, we derive new multimessenger bounds on the properties of their sources.
%We find that conventional scenarios suffer from a tension at $\gtrsim3\sigma$ level for a neutrino flux extending down to $10$~TeV, which is interpreted as evidence for a new class of sources that prohibit $\gamma$ rays from escaping them or have redshift evolution significantly different from the cosmic star formation rate.          
\end{abstract}

\maketitle

% body of paper here - Use proper section commands
% References should be done using the \cite, \ref, and \label commands
\section{\label{sec:intro}Introduction}
The origin of high-energy cosmic neutrinos has been one of the biggest enigmas in astroparticle physics since their discovery~\cite{Aartsen:2013bka,Aartsen:2013jdh,Aartsen:2015rwa,Halzen:2016gng}, and multimessenger relationships (i.e., among neutrinos, $\gamma$ rays, cosmic rays, and perhaps gravitational waves) have provided important clues to their sources~\cite{Ahlers:2017wkk,Ackermann:2019ows, Meszaros:2019xej,Murase:2019pef}. 
The fact that diffuse fluxes of PeV neutrinos, sub-TeV $\gamma$ rays, and ultrahigh-energy cosmic rays (UHECRs) are comparable suggests their physical connections~\cite{Murase:2016gly,Fang:2017zjf}. In particular, astrophysical neutrinos should originate from hadronuclear ($pp$) or photohadronic ($p\gamma$) interactions, in which the associated $\gamma$-ray production is unavoidable and the diffuse isotropic $\gamma$-ray background measured by the {\it Fermi} satellite~\cite{Ackermann:2014usa,Zechlin:2015wdz,TheFermi-LAT:2015ykq,Lisanti:2016jub} gives profound constraints on the candidate sources~\cite{Murase:2013rfa,Ahlers:2013xia}, and the importance of searching for neutrinos below 100~TeV has been emphasized~\cite{Murase:2013rfa}. 

The recent analyses of neutrino-induced showers and medium-energy starting events have revealed that the energy flux in the 10-100~TeV range is as large as $E_\nu^2\Phi_\nu\sim10^{-7}~{\rm GeV}~{\rm cm}^{-2}~{\rm s}^{-1}~{\rm sr}^{-1}$ with a steep spectral index of $s_{\rm ob}\gtrsim2.5-2.9$~\cite{Aartsen:2014muf,Aartsen:2015knd,Aartsen:2017mau,Aartsen:2020aqd}. This energy flux level is higher than the $\gtrsim0.1$~PeV neutrino flux obtained from upgoing muon neutrinos~\cite{Aartsen:2016xlq,Aartsen:2017mau} and exceeds many pre-discovery theoretical predictions~\cite{Waxman:1998yy}. 
The consistency with an isotropic distribution~\cite{Aartsen:2017mau} supports their extragalactic origin, even though a subdominant contribution may come from Galactic sources. This is further supported by the new shower data that are extended down to $\lesssim10$~TeV, which give $E_\nu^2\Phi_\nu=(1.66^{+0.25}_{-0.27})\times3\times10^{-8}~{\rm GeV}~{\rm cm}^{-2}~{\rm s}^{-1}~{\rm sr}^{-1}$ at 100~TeV (for the sum of all flavors) and $s_{\rm ob}=2.53\pm0.07$~\cite{Aartsen:2020aqd}.

It is critically important to identify and understand the sources of the medium-energy neutrinos in the 10-100~TeV range. 
First, the large neutrino flux compared to the extragalactic $\gamma$-ray background (EGB) flux~\cite{Ackermann:2014usa} is naturally explained by $\gamma$-ray hidden sources~\cite{Murase:2015xka}. If established, the IceCube data will enable us to utilize neutrinos as a unique probe of particle acceleration in dense environments. 
Second, the large neutrino flux implies that the energy generation rate density of high-energy neutrinos is significant as one of the nonthermal energy budgets in the Universe~\cite{Murase:2018utn}. Not many source candidates can satisfy the energy budget requirement, and possible candidate sources include the cores of active galactic nuclei (AGNs)~(see a review~\cite{Murase:2015ndr}) and choked-jet supernovae (SNe)~\cite{Meszaros:2001ms,Razzaque:2004yv,Ando:2005xi,Iocco:2007td,Murase:2013ffa}. Revealing the sources is also important for us to understand the multi-messenger connection among neutrinos, $\gamma$ rays and UHECRs. For example, the medium-energy neutrino flux cannot be explained by conventional $\gamma$-ray transparent sources such as galaxy clusters and starburst galaxies~\cite{Loeb:2006tw,Murase:2008yt,Kotera:2009ms}, so that a multi-component model may be required for the IceCube data from TeV to PeV energies~\cite{Kimura:2014jba,Chen:2014gxa,Murase:2015xka,Palladino:2016zoe,Chianese:2016smc,Chianese:2016kpu,Chianese:2017nwe,Palladino:2018evm}.

With the latest IceCube data in the 10-100~TeV range~\cite{Aartsen:2017mau,Aartsen:2020aqd}, and the EGB data from {\it Fermi-Lat}~\cite{Ackermann:2014usa}, this work provides the first quantitative constraints on the parameter space allowed by intent neutrino sources, from which GeV-TeV $\gamma$ rays escape. We show that the conventional $\gamma$-ray transparent scenario suffers from the $\gtrsim3\sigma$ tension with EGB data, which is regarded as evidence for hidden cosmic-ray accelerators or unknown Galactic sources.   

%%%%%%%%%%%%%%%%%%%%%
%%%%%%%%%%%%%%%%%%%%%
\section{\label{sec:g-nu}Modeling of $\nu$ and $\gamma$-ray Spectra}
%%%%%%%%%%%%%%%%%%%%%
%%%%%%%%%%%%%%%%%%%%%

Production of high-energy neutrinos in astrophysical sources requires hadronic processes creating $\pi^\pm$ and $K^\pm$ which subsequently decay to neutrinos; {\it e.g}, $\pi^+\to\mu^+\nu_\mu\to e^+\nu_\mu\bar{\nu}_\mu\nu_e$. The pions can be produced via interactions of accelerated protons with ambient protons ($pp$ scenarios) or photons ($p\gamma$ scenarios). The resulting neutrino flux has different characteristics in each scenario: while in the $pp$ scenario the neutrino spectrum extends to lower energies and increases with the decrease in energy, in the $p\gamma$ scenario a large fraction of produced neutrinos have energies larger than the threshold energy $\sim 4\times10^{-2}m_{\pi}m_p/\varepsilon_t\sim6\times10^6\,{\rm GeV}\, ({\rm eV}/\varepsilon_t)$ (where $\varepsilon_t$ is the energy of target photons), below which the neutrino spectrum drops rapidly.
For this work, we parametrize the neutrino spectrum in the $p\gamma$ scenario by introducing a break energy, $\varepsilon_{\rm br}$, where for $\varepsilon_\nu<\varepsilon_{\rm br}$ the spectrum hardens due to pion-decay kinematics, and we take it $\propto \varepsilon_\nu^{-s_l}$ with $s_l=0$~\cite{Murase:2015xka}. Theoretical calculations of $\varepsilon_{\rm br}$ require detailed knowledge on the source characteristics. We take the following energy spectrum:
\begin{equation}\label{eq:spec}
\varepsilon_\nu Q_{\varepsilon_\nu} \propto \begin{cases} 
\varepsilon_\nu^2 & \varepsilon_\nu < \varepsilon_{\rm br} \\
\varepsilon_\nu^{2-s_h} & \varepsilon_{\rm br}\leq \varepsilon_\nu \leq 10\,{\rm PeV} \\
0 & \varepsilon_\nu > 10\,{\rm PeV} 
\end{cases}~,
\end{equation}
where $\varepsilon_\nu Q_{\varepsilon_\nu} = n_s \varepsilon_\nu\frac{{\rm d}L_\nu}{{\rm d \varepsilon_\nu}}$ is the differential energy generation rate density of neutrinos with energy $\varepsilon_\nu$ for neutrino luminosity $L_\nu$ and the number density of the sources $n_s$. 
The neutrino flux is conservatively set to zero for $\varepsilon_\nu>10$~PeV since, so far, there is no observed neutrino flux at this energy range~\cite{Aartsen:2016ngq,Aartsen:2018vtx}.

We emphasize that the neutrino spectrum in Eq.~(\ref{eq:spec}) is the {\it minimal} assumption about the neutrino production in the source(s) that can accommodate the diffuse neutrino flux observed by IceCube. Extending the energy range either to lower energies, as in $pp$ scenario, or to higher energies, by increasing the assumed 10 PeV cutoff, increases the accompanying $\gamma$-ray flux. 

The energy flux observed at the Earth from a source at redshift $z$ is $\varepsilon_\nu ({\rm d}L_\nu/{\rm d}\varepsilon_\nu)|_{\varepsilon_\nu=(1+z)E_\nu}/(4\pi d_L^2)$, where $d_L$ is the luminosity distance, $H(z)$ is the $z$-dependent Hubble parameter and $E_\nu$ is the observed neutrino energy at Earth. Knowing the differential energy generation rate density of neutrinos at redshift $z=0$, which is $\varepsilon_\nu Q_{\varepsilon_\nu}$ in Eq.~(\ref{eq:spec}), and the dimensionless redshift evolution of the sources, $\mathcal{F}(z)$, which we take to be the cosmic start formation rate discussed in the Appendix, the all-flavor diffuse flux of neutrinos at the Earth from the distribution of sources is given by
\begin{equation}\label{eq:phinu}
E_\nu^2 \Phi_{\nu}^{\rm diff}=\frac{1}{4\pi}\int_0^\infty {\rm d}z\, \frac{{\rm d}\mathcal{V}_c}{{\rm d}z}\, \frac{\varepsilon_\nu Q_{\varepsilon_\nu}\mathcal{F}(z)}{4\pi d_L^2}~,
\end{equation}
where ${\rm d}\mathcal{V}_c/{\rm d}z = 4\pi[c/H(z)]d_L^2/{(1+z)}^2$ and $\mathcal{V}_c$ is the comoving volume.   

The $\gamma$-ray flux accompanied by the neutrino flux is calculated using the following argument: from isospin symmetry, not only $\pi^\pm$ but also $\pi^0$ have to be produced at the sources. The subsequent decay of $\pi^0$ to photons ($\pi^0\to2\gamma$) generates a $\gamma$-ray spectrum given by
\begin{equation}\label{eq:nug}
\varepsilon_\gamma Q_{\varepsilon_\gamma} = \frac{4}{3K} \left[ \varepsilon_\nu Q_{\varepsilon_\nu} \right]_{\varepsilon_\nu = \varepsilon_\gamma/2}~,
\end{equation} 
where $K\approx1$ for $p\gamma$ sources. The $\gamma$-ray flux should generally be larger than Eq.~(\ref{eq:nug}) because the charged pions can lose part of their energies before the decay, adiabatically or radiatively, and also ambient electrons and positrons can enhance the $\gamma$-ray production via cascades inside the sources.

The calculation of the $\gamma$-ray flux at Earth is more complicated than the neutrino flux. Even for the most conservative setup, in which the sources are optically thin to $\gamma$ rays, the Universe is opaque to $\gamma$ rays with energy $\gtrsim 1$~TeV that propagate distances $z\gtrsim10^{-2}$, due to the absorption by pair production on the Cosmic Microwave Background (CMB) and Extragalactic Background Light (EBL) photons~\cite{Dominguez:2010bv}.
At $\gtrsim1$~PeV energies, this absorption is significant even at the Galactic scale~\cite{Ahlers:2013xia,Esmaili:2015xpa}. However, the pairs produced in the pair production process inverse-Compton scatter off the CMB and EBL, creating new $\gamma$ rays at slightly lower energies than the original $\gamma$ rays. These successive processes initiate an electromagnetic cascade which ceases at the pair-production threshold $\sim m_e^2/\varepsilon_{\rm t}$. 
For the CMB, this cutoff appears at $\sim100$~TeV, while for the EBL it is $\sim100$~GeV. Thus, although the Universe is opaque to high-energy photons, the initial high-energy $\gamma$-ray flux will be redistributed in the GeV-TeV range due to the electromagnetic cascade. 
Although the approximate spectrum can be calculated analytically~\cite{Berezinsky:1975zz,Ginzburg:1990sk}, the exact energy dependence of the flux needs numerical calculations taking into account the $z$-dependence of the EBL and CMB. 
In this work, we use the public code $\gamma$-Cascade for this purpose~\cite{Blanco:2019oa}, which agrees well with results of the previous literature~\cite{Murase:2013rfa}.

%%%%%%%%%%%%%%%%%%
%%%%%%%%%%%%%%%%%%
\section{\label{sec:ana}Multimessenger Analyses}       
%%%%%%%%%%%%%%%%%%
%%%%%%%%%%%%%%%%%%   

The diffuse and isotropic $\gamma$-ray flux arising from cascades induced by high-energy photons in the intergalactic space contributes to the EGB. A conservative limit on the neutrino sources has been derived in Ref.~\cite{Murase:2013rfa} by requiring that the resulting flux should not overshoot the isotropic diffuse $\gamma$-ray background (IGRB) part of the then measured EGB~\cite{Abdo:2010nz} (which was extending to $\sim100$~GeV) at any energy. In our analysis we consider the latest measured whole EGB data~\cite{Ackermann:2014usa} which is extending to 820~GeV.
However, there are other contributions to the EGB originating from populations of unresolved sources at low energies, \textit{i.e.,} $\lesssim 1$~TeV, which should be taken into account, such as the guaranteed contributions from jetted AGNs (including blazars and radio galaxies), star-forming galaxies and cosmogenic $\gamma$ rays~\cite{Fornasa:2015qua}. 
One approach is to subtract such point-source contributions from the EGB and use the remaining flux to set limits on any additional diffuse $\gamma$-ray contribution, including the cascaded flux that we are interested in. A more appropriate approach is to perform a $\chi^2$ analysis by taking into account the various contributions.

In the following we describe two different analyses performed in this work: 
\begin{enumerate}
\item[\textbf{A}.] \textbf{$\chi^2$ analysis}: The contribution of blazars to the EGB, including BL Lac objects and Flat Spectral Radio Quasars (FSRQs), has been calculated in Ref.~\cite{Ajello:2015mfa}. We use their luminosity-dependent density evolution (LDDE) model for the luminosity function of the blazars. The emissions from star-forming galaxies~\cite{Ackermann:2012vca} and radio galaxies~\cite{Inoue:2011bm} have also been taken into account. Using these contributions and the EGB data we set a limit on any extra contribution to the EGB by defining the following $\chi^2$ function:
\begin{equation}\label{eq:chi2}
\chi^2 = \min_\mathcal{A} \left[ \sum_i \frac{\left(F_{i,{\rm EGB}}-\mathcal{A}F_{i,{\rm a}}-F_{i,{\rm cas}}\right)^2}{\sigma_i^2} + \frac{\left(\mathcal{A}-1\right)^2}{\sigma_\mathcal{A}^2}\right]~,
\end{equation}
where $F_{i,{\rm EGB}}$, $F_{i,{\rm a}}$ and $F_{i,{\rm cas}}$ are, respectively, the observed EGB flux, the astrophysical contribution (blazars, star-forming galaxies and radio galaxies) and the cascaded flux contribution to the $i$-th energy bin. Also $\sigma_i$ is the uncertainty on EGB flux and the last term is the pull-term, taking into account the normalization uncertainty of the astrophysical contribution, given by $\sigma_\mathcal{A}\approx35\%$~\cite{Ajello:2015mfa}.

\item[\textbf{B}.] \textbf{Integrated flux above $50$~GeV}: It has been shown in Ref.~\cite{TheFermi-LAT:2015ykq} that $86^{+16}_{-14}\%$ of the total EGB above $50$~GeV can be accounted for by the contribution from the sources in the 2FHL catalog, mainly consisting of blazars. The total EGB integrated flux above $50$~GeV is $J_{>50~{\rm GeV}}^{\rm EGB} = 2.4\times10^{-9}~{\rm ph}/{\rm cm}^2/{\rm s}/{\rm sr}$. So, by requiring 
\begin{equation}\label{eq:50}
\int_{50~{\rm GeV}}^{820~{\rm GeV}} \Phi^{\rm cas}_\gamma~{\rm d}E_\gamma < (1-q)J_{>50~{\rm GeV}}^{\rm EGB}~,
\end{equation}
we can derive limits on the cascaded $\gamma$-ray flux $\Phi^{\rm cas}_\gamma$. In the above relation, $q$ is the percentage of total EGB intensity (above 50 GeV) which can be explained by the blazars, with central value $q=86\%$.    
       
\end{enumerate}

Method \textbf{B} is an independent analysis which is not sensitive to the spectral shape of the cascaded flux as in method \textbf{A}. Also, the majority of sources in the 2FHL catalog are blazars (and among those, $74\%$ are BL Lac objects). This means that the constraints derived from method \textbf{B} are very conservative and based on the contribution of a single source population to the EGB. Although the principal result of this Letter comes from the method \textbf{A}, we perform the analysis of method \textbf{B} as a sanity check.     

The cascaded $\gamma$-ray flux from the distribution of sources responsible for the neutrino flux observed in IceCube depends on $s_h$ and $\varepsilon_{\rm br}$ via Eq.~(\ref{eq:spec}) (through Eq.~(\ref{eq:nug})). Using the EGB data, we can derive constraints on the $s_h$ and $\varepsilon_{\rm br}$ parameters, or equivalently, on $s_h$ and $E_{\rm br}$, as well as on the normalization of the corresponding neutrino flux. Here the $E_{\rm br}$ is the redshifted observed energy break at Earth (see Appendix).

%%%%%%%%%%%%%%%%%%
%%%%%%%%%%%%%%%%%%
\section{\label{sec:res}Results}       
%%%%%%%%%%%%%%%%%%
%%%%%%%%%%%%%%%%%% 

\begin{figure}[h!]
\includegraphics[width=0.48\textwidth]{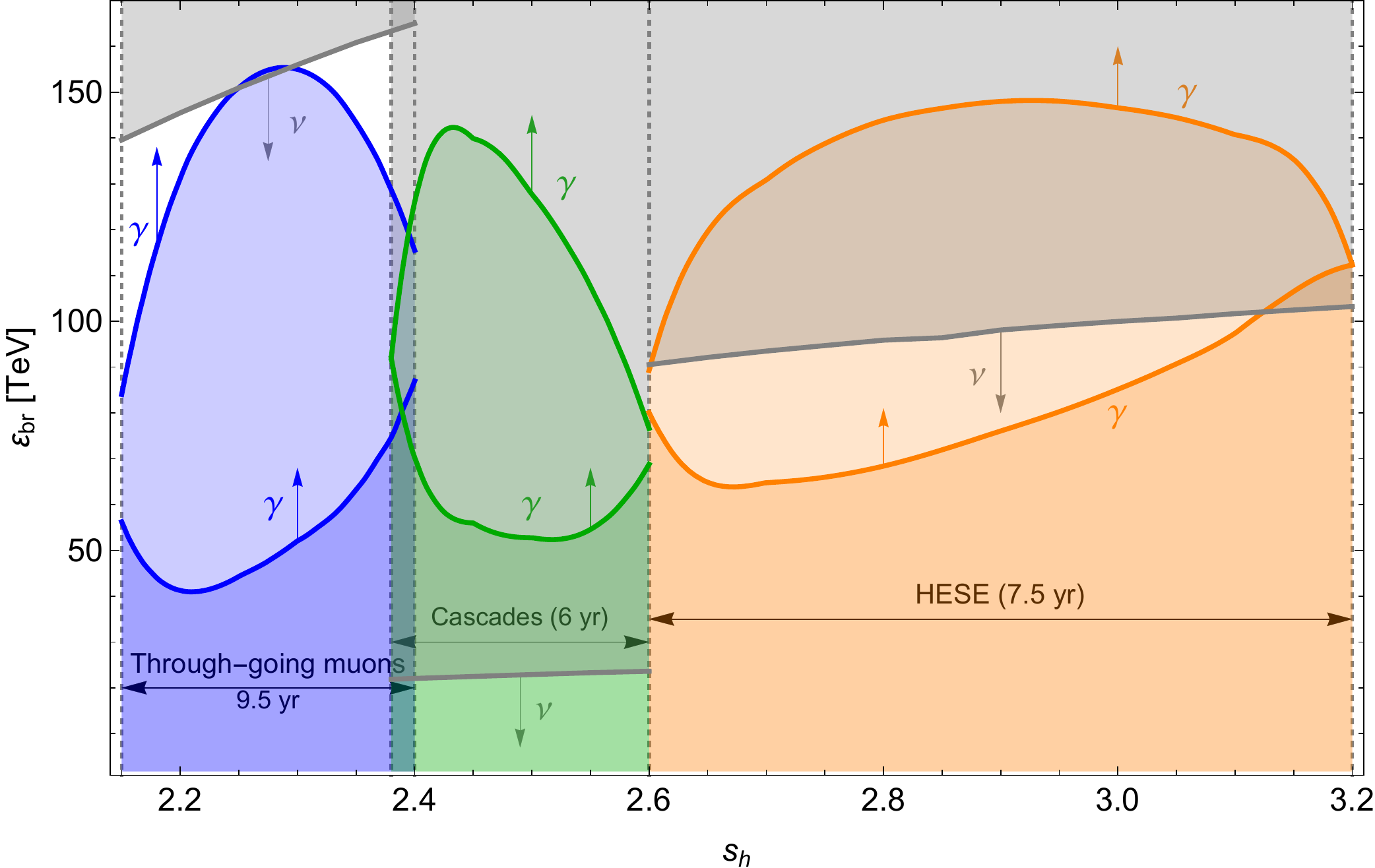}
\caption{\label{fig:chi2}The 90\% C.L. constraints on $\varepsilon_{\rm br}$ vs. $s_h$ for the three data sets of IceCube, from method \textbf{A} of analyzing the EGB data. The gray- and color-shaded regions show the exclusions based on IceCube data and EGB data, respectively (the arrows point toward allowed regions). For each color, the upper and lower curves respectively correspond to the maximum and minimum allowed flux normalizations, $\Phi_{\rm astro}$, at $1\sigma$ reported by IceCube.}
\end{figure}

In analysis method \textbf{A}, constraints in the $(s_h,\varepsilon_{\rm br})$ plane, shown in Figure~\ref{fig:chi2}, are derived by defining $\Delta\chi^2=\chi^2-\chi^2_{\rm min}$, where $\chi^2_{\rm min}$ is the minimum value of $\chi^2$ in Eq.~(\ref{eq:chi2}) without the pull-term (free $\mathcal{A}$) and a free normalization for the cascaded $\gamma$-ray flux. The $s_h$ range for each IceCube data set, see Appendix, is depicted and the gray shaded regions show the excluded $\varepsilon_{\rm br}$ from IceCube data (by translating the $E_{\rm br}$ to $\varepsilon_{\rm br}$ for each data set). The color-shaded regions show the excluded $\varepsilon_{\rm br}$ values from the EGB data at 90\% C.L. limits derived from the condition $\Delta\chi^2<4.61$ (for 2 d.o.f.). For each color (corresponding to a different IceCube analysis), the upper and lower curves correspond, respectively, to the highest and lowest IceCube allowed normalizations, $\Phi_{\rm astro}$ defined in Eq.~(\ref{eq:icflux}) in the Appendix, at $1\sigma$ (shown in Figure~\ref{fig:phi}). Clearly, from Figure~\ref{fig:chi2}, the HESE and through-going $\nu_\mu$-track data sets of IceCube are compatible with the EGB data, while the measured neutrino flux in the cascade data set leads to a diffuse $\gamma$-ray flux that is incompatible with EGB data.

%%%%%%%         FIGURE   1        %%%%%%%
%%%%%%%%%%%%%%%%%%%%%%%%%
%%%%%%%%%%%%%%%%%%%%%%%%%
%%%%%%%%%%%%%%%%%%%%%%%%%

To quantify the tension in Figure~\ref{fig:chi2}, using method \textbf{A}, we derive constraints in the $(s_h,\Phi_{\rm astro})$ plane for fixed values of $E_{\rm br}$. The color-shaded regions in Figure~\ref{fig:phi} show the allowed regions in each IceCube data set in the $(s_h,\Phi_{\rm astro})$ plane. The solid curves show the limits, at $2\sigma$ C.L., from method \textbf{A} of analyzing the EGB data for the depicted $E_{\rm br}$ values, where the arrows point toward the allowed regions. We can see that having astrophysical neutrinos down to $\sim10$~TeV, as the 6-year cascade data set indicates~\cite{Aartsen:2020aqd}, leads to a tension with the EGB data. As in Figure~\ref{fig:chi2}, the HESE and through-going $\nu_\mu$-track analyses rely on the data above $\sim60$~TeV and $\sim120$~TeV, respectively, so they are compatible with the EGB data. Both the 4-year~\cite{Niederhausen:2017mjk} and 6-year~\cite{Aartsen:2020aqd} cascade data sets are essential for the tension. From Figure~\ref{fig:phi}, we can also conclude that extending the astrophysical neutrino flux to energies $\lesssim20$~TeV results in tensions with the EGB data for all the three sets of IceCube data. The present shower data with $E_{\rm br}\approx10$~TeV is in tension with the EGB data at $\gtrsim3\sigma$ C.L., whereas for $E_{\rm br}\approx1$~TeV, it grows to $\approx5\sigma$. The statistical significance of this tension increases in a more realistic setup.  

%%%%%%%         FIGURE   2        %%%%%%%
%%%%%%%%%%%%%%%%%%%%%%%%%
\begin{figure}[h!]
\includegraphics[width=0.48\textwidth]{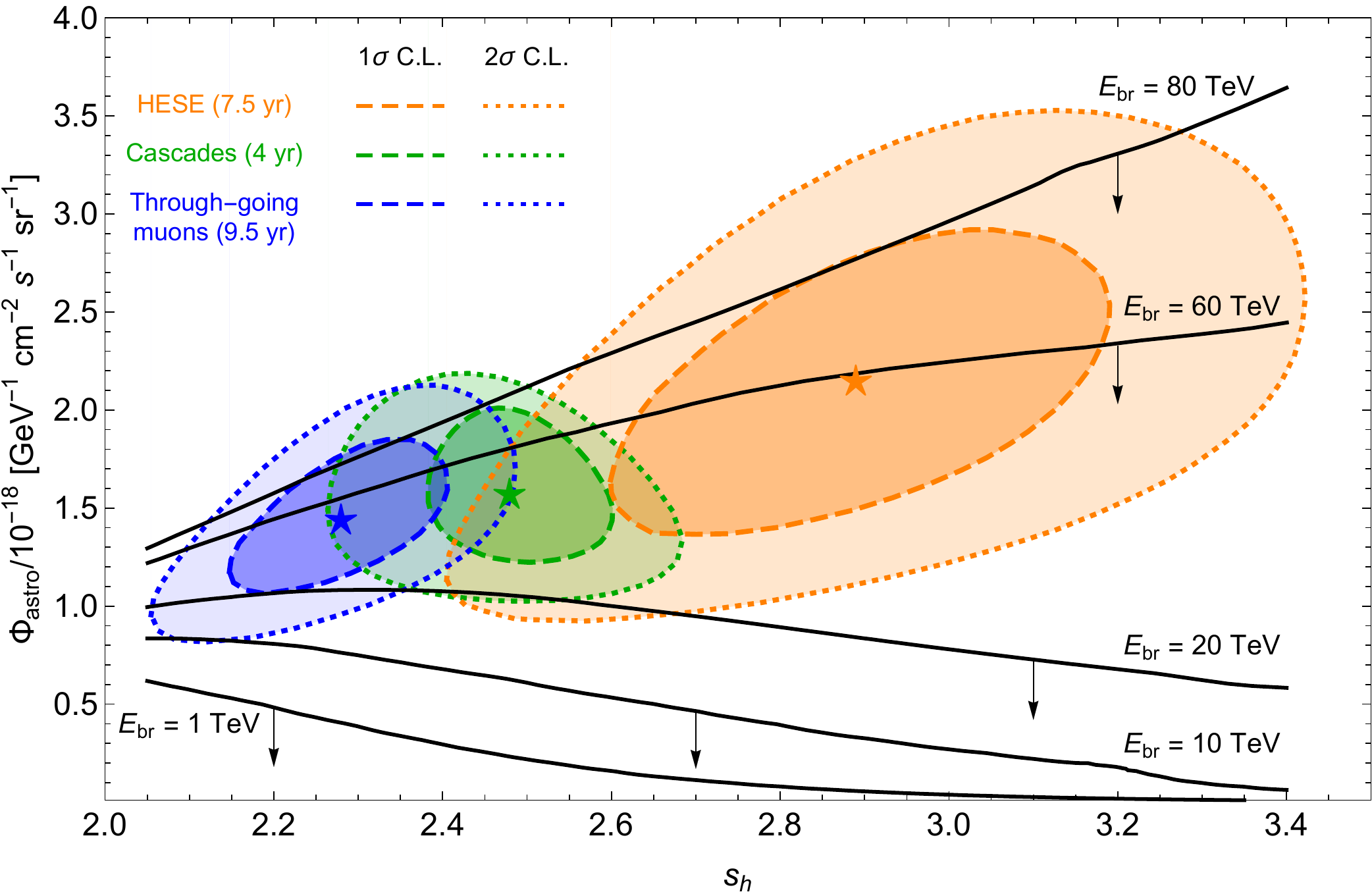}
\caption{\label{fig:phi}
Constraints in the $(s_h,\Phi_{\rm astro})$ plane from method \textbf{A} of analyzing the EGB data. The solid black curves depict the allowed regions, for fixed $E_{\rm br}$, from EGB data. The green shaded regions show the allowed regions for the 4-year cascade events~\cite{Niederhausen:2017mjk} which are similar to the 6-year cascade~\cite{Aartsen:2020aqd} allowed regions.}
\end{figure}
%%%%%%%%%%%%%%%%%%%%%%%%%
%%%%%%%%%%%%%%%%%%%%%%%%%

\begin{figure}[h!]
\includegraphics[width=0.48\textwidth]{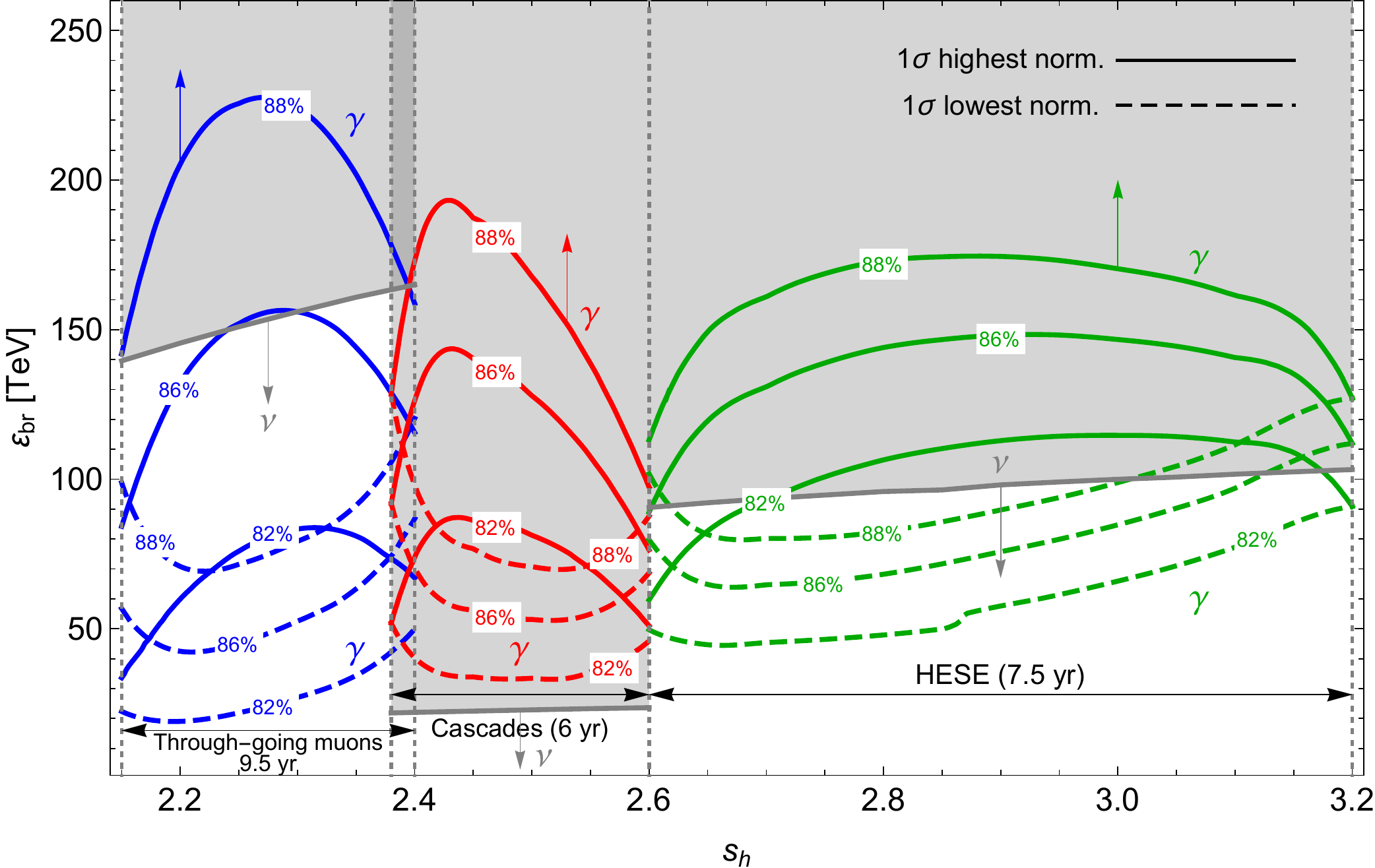}
\caption{\label{fig:int}Constraints on $\varepsilon_{\rm br}$ vs. $s_h$, as in Fig.~\ref{fig:chi2}, this time from method \textbf{B}. The labels on the curves show the percentage of EGB flux above 50~GeV that can be accounted by the blazars in the 2FHL catalog (the $q$ value in Eq.~(\ref{eq:50})).}
\end{figure}

As an independent analysis, Figure~\ref{fig:int} shows the results based on method \textbf{B}. The solid (dashed) curves correspond to the highest (lowest) allowed normalization of astrophysical neutrinos at $1\sigma$ level. The label on each curve shows the $q$ value in Eq.~(\ref{eq:50}). Consistent with method \textbf{A}, Figure~\ref{fig:int} shows the tension between the IceCube cascade data set and the EGB data for $q\gtrsim80\%$. Obviously, method \textbf{B} is less constraining since the analysis is based on just the \textit{integrated} flux of EGB above 50~GeV and is independent of the spectral shape of the cascaded flux, which in fact is important at $\sim 100$~GeV.   

The redshift evolution slightly affects the tension quantitatively but not qualitatively and the conclusions remain the same for redshift evolution of the most of the source classes including galaxy clusters, star-forming galaxies, and AGNs~\cite{Murase:2013rfa}.    

%%%%%%%         FIGURE   3        %%%%%%%
%%%%%%%%%%%%%%%%%%%%%%%%%
%%%%%%%%%%%%%%%%%%%%%%%%%
%%%%%%%%%%%%%%%%%%%%%%%%%

%%%%%%%%%%%%%%%%%%
%%%%%%%%%%%%%%%%%%
\section{\label{sec:con}Summary and Discussions}       
%%%%%%%%%%%%%%%%%%
%%%%%%%%%%%%%%%%%% 

The neutrino flux observed in IceCube should be accompanied by the $\gamma$-ray flux, which provides a powerful diagnostic in the search for their possible sources. Assuming a minimal model for high-energy cosmic neutrinos, for the first time, we showed that the new IceCube data extended down to $\sim10$~TeV leads to $\gtrsim3\sigma$ tension with the EGB data from {\it Fermi-LAT}. The significance of tension increases to $\sim5\sigma$ for astrophysical neutrino $\sim1$~TeV. We stress that the derived limits and reported tension are based on very conservative assumptions. The tension is $\approx3\sigma$ for a break energy of $E_{\rm br}\approx10$~TeV, and larger for more realistic setups. First, the neutrino spectrum is modified by the cooling of mesons and muons, which yields a larger ratio of $\gamma$ rays to neutrinos. Second, additional $\gamma$ rays must be produced by the Bethe-Heitler process; for example, these Bethe-Heitler-induced $\gamma$ rays are dominant in the AGN core scenario~\cite{Murase:2019vdl}. Third, $\gamma$ rays should also be produced by leptonic processes which do not produce any neutrinos. GeV-TeV $\gamma$ rays of blazars are conventionally explained by the leptonic components. 

The reported tension suggests an additional population of the sources, which are different from conventional cosmic-ray reservoirs.  
Hidden ($\gamma$-ray opaque) cosmic-ray accelerators are among the promising sources of the medium-energy IceCube neutrinos. Candidate classes include choked GRB jets~\cite{Murase:2013ffa,Tamborra:2015fzv,Senno:2015tsn}, AGN cores~\cite{Stecker:2013fxa,Kalashev:2015cma,Murase:2019vdl,Inoue:2019fil,Kimura:2019yjo}, and MeV blazars~\cite{Murase:2015xka}. 
Alternatively, high-redshift source population that do not exist in the local universe can alleviate the tension. For example, with the redshift evolution of POP-III stars, the EBL cutoff can be down to 10~GeV energies~\cite{Xiao:2016rvd}. 
Finally, in principle, Galactic sources that lead to quasi-isotropic emission, such as the Galactic halo~\cite{Ahlers:2013xia,Taylor:2014hya,Liu:2018gyj}, may give a significant contribution. Although the 10-100~TeV neutrinos come from both hemispheres and there is a tension with some of the upper limits from air-shower experiments~\cite{Ahlers:2013xia,Murase:2015xka}, further multimessenger studies are necessary~\cite{Neronov:2020wir}. In fact, the claimed upturn in the IGRB~\cite{Neronov:2018ibl} can support such Galactic halo scenarios.   
Our results also impact nonastrophysical scenarios that explain the shower data with physics beyond the Standard Model (BSM) (see reviews~\cite{Ahlers:2018mkf,Ackermann:2019cxh}). 
For example, decaying dark matter has been invoked as an interpretation of the IceCube data~\cite{Esmaili:2013gha,Feldstein:2013kka} (see Refs.~\cite{Bhattacharya:2017jaw,Bhattacharya:2019ucd,Chianese:2019yu} for recent analyses). Final states involving quarks, charged leptons and gauge bosons are accompanied by a comparable $\gamma$-ray flux~\cite{Murase:2015gea,Esmaili:2015xpa}, which gives strong constraints especially for models explaining the medium-energy neutrino data~\cite{Cohen:2016uyg,Hiroshima:2017hmy,Chianese:2018ijk,Ishiwata:2019aet}. 
Other BSM explanations, such as neutrino decay~\cite{Denton:2018aml}, increase the ratio of $\gamma$ rays to neutrinos, which further strengthens the results of this work~\cite{Bustamante:2016ciw}.  

Further observations of the medium-energy range (by more efficient rejection of background events) to lower energies is of crucial importance. IceCube-Gen2 will give us more statistics, but the threshold energy should not be far from $\sim10$~TeV. KM3NeT~\cite{Adrian-Martinez:2016fdl} will be able to give us information on the northern sky, which is complementary, and the detection of showers with a better angular resolution will be particularly useful. In addition, stacking searches with source catalogues at different wavelengths are strongly encouraged.  
Intriguingly, a hidden cosmic-ray accelerator with a steep neutrino spectrum is independently indicated from the recent $\sim3\sigma$ observation of NGC 1068~\cite{Aartsen:2019fau,Murase:2019pef}. 
Searching for lower-energy $\gamma$-ray counterparts in the MeV energy range will also be important.

% If you have acknowledgments, this puts in the proper section head.
\begin{acknowledgments}
We thank Ali Kheirandish and Pasquale Serpico for useful discussions and comments.   
This work has been supported by the Alfred P. Sloan Foundation and NSF Grant No.~AST-1908689 (K.M.). A.~E. thanks the partial support received by the CNPq fellowship No. 310052/2016-5. A. C. thanks the support received by the FAPERJ scholarship No. E-26/201.794/2019.
\end{acknowledgments}

\bibliography{kmurase}

%\newpage
\clearpage
\appendix

\section{Redshift evolution\label{app:red}}

\begin{figure}[h!]
\includegraphics[width=0.49\textwidth]{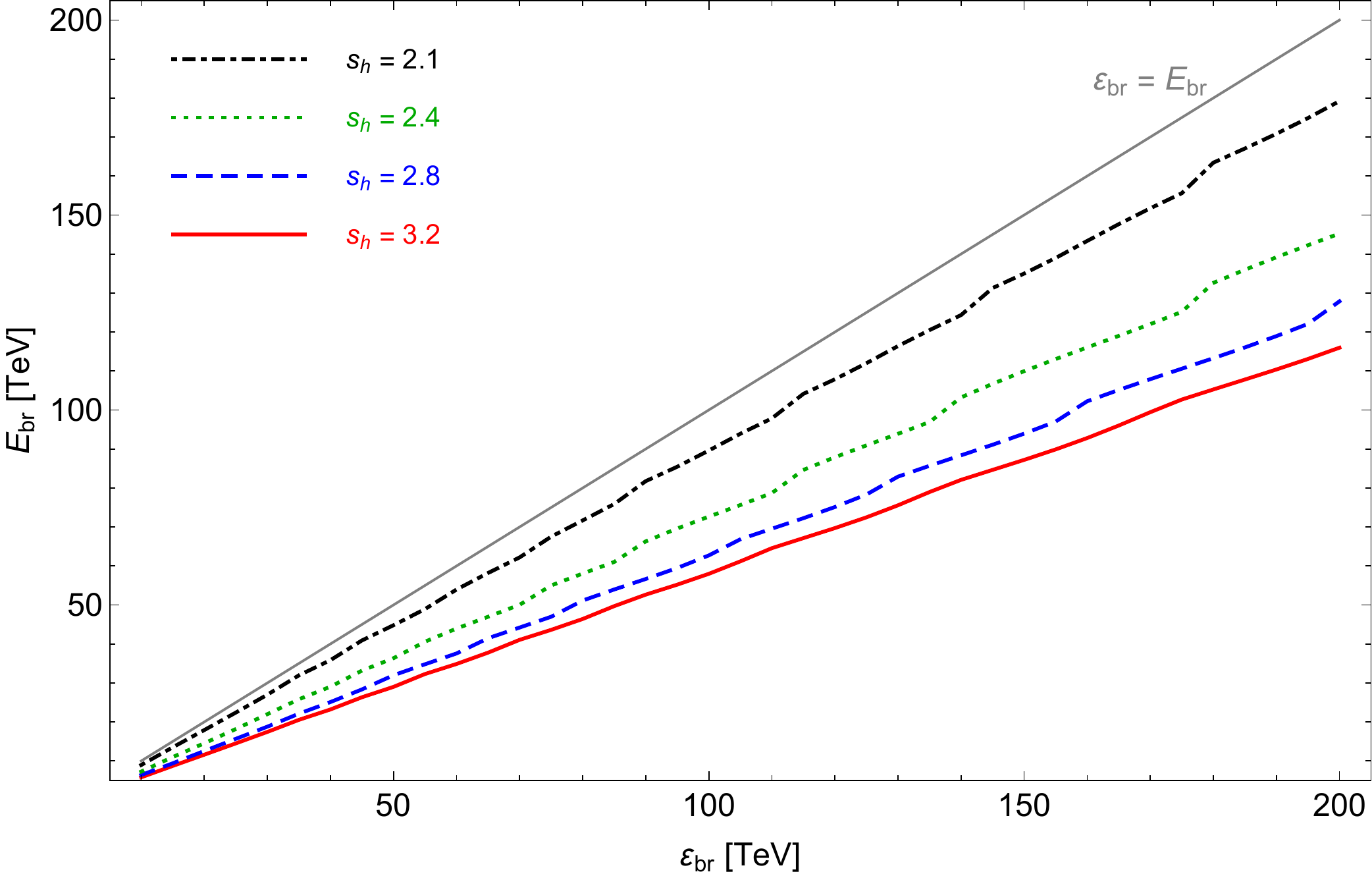}
\caption{\label{fig:ebr}The red-shifted energy break at Earth, $E_{\rm br}$, in terms of the $\varepsilon_{\rm br}$ at the sources, calculated for SFR evolution. The diagonal gray line shows $\varepsilon_{\rm br}=E_{\rm br}$.}
\end{figure}

For $\mathcal{F}(z)$ in Eq.~(\ref{eq:phinu}) and the corresponding computations of the $\gamma$-ray flux, we use the cosmic star formation rate (SFR)~\cite{Hopkins:2006bw,Yuksel:2008cu} given by
\begin{equation}\label{eq:SFR}
\mathcal{F}(z) =\left[(1+z)^{a\eta}+\left(\frac{1+z}{B}\right)^{b\eta}+\left(\frac{1+z}{C}\right)^{c\eta} \right]^{1/\eta}~,
\end{equation}
where $a=3.4$, $b=-0.3$ and $c=-3.5$. The constants $B\simeq 5000$ and $C\simeq 9$ correspond to breaks at $z\simeq1$ and $z\simeq4$, respectively, and $\eta=-10$ smooths the transition between the breaks. The normalization in Eq.~(\ref{eq:spec}) is fixed by the observed IceCube neutrino flux. Notice that the abrupt break in the injection spectrum at 10 PeV is smoothed at the Earth due to cosmological redshift. The same effect causes the position of the break in the energy spectrum at the Earth, $E_{\rm br}$, to be shifted with respect to the energy break at the source, $\varepsilon_{\rm br}$. In our case, this shift depends only on the spectral index $s_h$ and on our choice of adopting the SFR evolution. Figure~\ref{fig:ebr} shows $E_{\rm br}$ in terms of $\varepsilon_{\rm br}$ for various $s_h$ values, where the diagonal gray line depicts $E_{\rm br}=\varepsilon_{\rm br}$, that is, no redshift. We can see that an increase in $s_h$ results in a decrease in the ratio $E_{\rm br}/\varepsilon_{\rm br}$, and for $s_h\simeq3$, it reaches approximately 50\%.

%%%%%%%         FIGURE   4        %%%%%%%
%%%%%%%%%%%%%%%%%%%%%%%%%
%%%%%%%%%%%%%%%%%%%%%%%%%
%%%%%%%%%%%%%%%%%%%%%%%%%

%%%%%%%%%%%%%%%%%%
%%%%%%%%%%%%%%%%%%
\section{\label{sec:data}$\nu$ and $\gamma$ data sets}       
%%%%%%%%%%%%%%%%%%
%%%%%%%%%%%%%%%%%%

The astrophysical neutrino flux has been measured by IceCube in several channels. The channels can be characterized by the event topology, either cascade or $\nu_\mu$-track events, and the location of neutrino-nucleus vertex, which can be either inside or outside the fiducial volume of IceCube, leading to \textit{starting} or \textit{through-going} $\nu_\mu$-track events, respectively. The measured differential flux from the data in each channel can be parametrized by (in units of $[{\rm GeV}^{-1}{\rm cm}^{-2}{\rm s}^{-1}{\rm sr}^{-1}]$) 
\begin{equation}\label{eq:icflux}
\Phi_\nu = 10^{-18}\cdot \Phi_{\rm astro} \left( \frac{E_\nu}{100~{\rm TeV}}\right)^{-s_{\rm ob}}~,
\end{equation}
where $\Phi_{\rm astro}$ and $s_{\rm ob}$ are the (observed) normalization and energy index of the flux. Since the background events for each channel are different, the minimum observed energy, or threshold energy $E_{\rm thr}$, which depends on the efficiency of background rejection at low energies, varies among the data sets. 

We consider the following three data sets: {\it i}) $7.5$-years of High Energy Starting Events (HESE) over the full-sky~\cite{Schneider:2019ayi}, consisting of both cascade and $\nu_\mu$-track events with the interaction vertex inside the fiducial volume of IceCube and with the threshold energy $E_{\rm thr}=60$~TeV. The reported all-flavor normalization and energy index are ($1\sigma$ error) $\Phi_{\rm astro}=6.45^{+1.46}_{-0.46}$ and $s_{\rm ob}=2.89^{+0.2}_{-0.19}$. \textit{ii}) 6-years cascade events~\cite{Aartsen:2020aqd} over the entire sky with $E_{\rm thr}=16$~TeV, one-flavor normalization $\Phi_{\rm astro}= 1.66^{+0.25}_{-0.27}$ and energy index $s_{\rm ob}=2.53\pm0.07$. The precedent 4-year cascade data set~\cite{Niederhausen:2017mjk} has almost the same normalization and index. \textit{iii}) $9.5$-years of through-going $\nu_\mu$-track events over the northern hemisphere~\cite{Stettner:2019tok} with $E_{\rm thr}=119$~TeV, one-flavor normalization $\Phi_{\rm astro}=1.44^{+0.25}_{-0.24}$ and energy index $s_{\rm ob}=2.28^{+0.08}_{-0.09}$. All the reported normalization and energy index values in the three data sets come from single power-law fits to data. For all data sets, a broken power-law fit also has been performed showing no preference over the single power-law fit.

The $\gamma$-ray data set consists of the Extragalactic $\gamma$-ray Background (EGB) measured by the Large Area Telescope (LAT) on board the Fermi Gamma-ray Space Telescope (Fermi)~\cite{Ackermann:2014usa}. The EGB is the sum of contributions from all the extragalactic $\gamma$-ray sources, including individual sources (faint and unresolved sources) and diffuse ones such as the Galactic foreground and (possible) contributions from electromagnetic cascades and dark matter annihilation/decay. The latest EGB data set covers the energy range $100$~MeV to $820$~GeV.

\end{document}